\documentclass[preprint,tightenlines]{revtex4-1}

\usepackage{amsmath,amssymb}
\usepackage{bm}
\usepackage[dvipdfmx]{graphicx}
\usepackage{ascmac} 
\usepackage{fancybox}
\usepackage{float}
\usepackage{enumerate}
\usepackage{color}
\usepackage{amsthm}
\usepackage{mathrsfs}
\allowdisplaybreaks[1]
\usepackage{comment}
\newtheorem*{th.}{Theorem}

\newcommand{\tr}{{\rm Tr}}

\newcommand{\ket}[1]{|#1\rangle}
\newcommand{\bra}[1]{\langle #1|}

\begin{document}
\title{Controllable non-Markovianity in phase relaxation}
\date{\today}
\author{Shingo Kukita$^{1,2)}$}
\email{toranojoh@shu.edu.cn}
\author{Yasushi Kondo$^{2)}$}
\email{ykondo@kindai.ac.jp}
\author{Mikio Nakahara$^{1,3)}$}
\email{nakahara@shu.edu.cn}
\affiliation{$^{1)}$Department of Mathematics, 
Shanghai University, Shanghai 200444, China}
\affiliation{$^{2)}$Department of Physics, 
Kindai University, Higashi-Osaka 577-8502, Japan}
\affiliation{$^{3)}$Research Institute for Science and Technology, Kindai University, Higashi-Osaka 577-8502, Japan}
\begin{abstract}
Recently remarkable progress in quantum technology has been witnessed. 
In view of this it is important to investigate an open quantum system
as a model of such quantum devices.
Quantum devices often require extreme conditions such as very low temperature 
for the devices to operate. 
Dynamics can be non-Markovian in such a situation in contrast with Markovian
dynamics in high temperature regime. 
This observation necessitates us to investigate a non-Markovian 
open quantum system,
both theoretically and experimentally.
In this paper, we report two important results:
1) Exact solution of a simple but non-trivial theoretical model 
and 2) demonstration of this model by NMR experiments, 
where non-Markovianity is continuously controllable.
We observe qualitative agreement between theory and experiment.
\end{abstract}
\maketitle
\newpage
\section{Introduction}
\label{sec:1}
Quantum resources provide us with novel protocols 
in several fields in particular in quantum information processing, such as 
quantum communication, quantum computing
and quantum sensing \cite{Nielsen2000}.
Many of such protocols have already been demonstrated in actual physical 
systems thanks to the advance of quantum technology.
Since quantum devices suffer from environmental noise, it is important 
to investigate open quantum systems \cite{Weiss}.
Quantum devices are often cooled down to very low temperature to make
the devices work. In such situations, the system dynamics often shows 
non-Markovian behaviour~\cite{GORINI1989357,RevModPhys.88.021002,
RevModPhys.89.015001,C4CP04922E,PhysRevA.86.012115}
while Markovian one is observed commonly in higher temperature regime.
Therefore, it is necessary for us to investigate open quantum systems 
in various environments theoretically~\cite{PhysRevA.55.2290,
PhysRevLett.120.030402,PhysRevA.87.040103}
and experimentally~\cite{Chiuri:2012aa,PhysRevA.91.012122,
PhysRevLett.121.060401,PhysRevLett.124.210502}.
It is, however, generally difficult to experimentally control 
non-Markovianity of a system dynamics.

Recently, a simple model that showed
time-homogeneous, time-inhomogeneous Markovian relaxations and
non-Markovian relaxations was proposed in Ref.~\cite{Ho_2019}. 
The system considered was composed of three subsystems,
namely, System~I (principal system), Markovian environment and System~II
inserted between System~I and the environment. 
They analysed the dynamics of this system 
by solving the Gorini-Kossakowski-Lindblad-Sudarshan (GKLS) master equation
\cite{Lindblad1976,GKS} 
analytically. They found that
the characteristics of relaxation
of System~I was controlled by tuning parameters of the environment as well as
coupling/decoupling System~II with System~I. 
Moreover, they experimentally demonstrated
the theoretical results with star-topology molecules in isotropic liquids
by using NMR. 
Through their study, it was found that System II worked
as a temporal storage of quantum information that was stored 
in System I and was dissipating into the Markovian environment.

The coupling between System~I and System~II in Ref.~\cite{Ho_2019}
was simply turned on and
off by using an NMR technique called decoupling \cite{Levitt2008}.
It is the purpose of this paper to further extend the model discussed 
in Ref.~\cite{Ho_2019}
by controlling the coupling strength between System~I and System~II.
We analyze this model by solving the GKLS master 
equation~\cite{Lindblad1976,GKS} 
analytically under some reasonable 
assumptions~\cite{semigroupMEq,Viola2013,Brito_2015} 
and compare the theoretical results with those obtained from 
liquid state NMR experiments.
It turns out that our model continuously interpolates between Markovian 
regime and non-Markovian regime.

The rest of this paper is organized as follows.
In Sec.~\ref{sec:2}, we introduce our theoretical model that is made of 
System~I, System~II and environment. It is shown that
non-Markovianity of the principal system dynamics is controlled 
by adjusting an external field applied to System~II.
The dynamics is studied by solving the GKLS equation
analytically.
We conducted NMR experiments, in which our theoretical model was 
implemented, and
compare theoretical predictions with experimental results in Sec.~\ref{sec:3}.
We introduce  a quantitative measure of non-Markovianity 
in our dynamics in Sec.~\ref{sec:2} and compare the theoretical prediction of this measure 
with experimental results in Sec.~\ref{sec:3}. 
Section \ref{sec:4} is devoted to conclusion and discussion. 
Details of some derivations are given in Appendix.

\section{ENGINEERED ENVIRONMENT: THEORY}
\label{sec:2}

It is well known that a quantum system relaxes exponentially
if it interacts with an environment that has an infinitesimally short memory.
This process is called ``Markovian''.
On the other hand, the relaxation is non-exponential when the system 
interacts with an environment with a long-time memory. In this case, 
information of the system temporarily stays
in the surrounding environment before it totally dissipates. 
We call this process ``non-Markovian''.
There are many studies on Markovian and non-Markovian dynamics;
in particular, non-Markovian dynamics is currently attracting 
much attention \cite{Weiss,GORINI1989357,RevModPhys.88.021002,
RevModPhys.89.015001,C4CP04922E,PhysRevA.86.012115,Vacchini_2011,
PhysRevA.73.012111}.
Non-Markovian dynamics often manifests itself in low 
temperature~\cite{GORINI1989357,RevModPhys.88.021002,
RevModPhys.89.015001,C4CP04922E,PhysRevA.86.012115}, 
small size environment, and/or strong coupling regime, for example .

\begin{figure}[h]
\begin{minipage}[t]{0.48\hsize}
\begin{center}
\includegraphics[width=60mm]{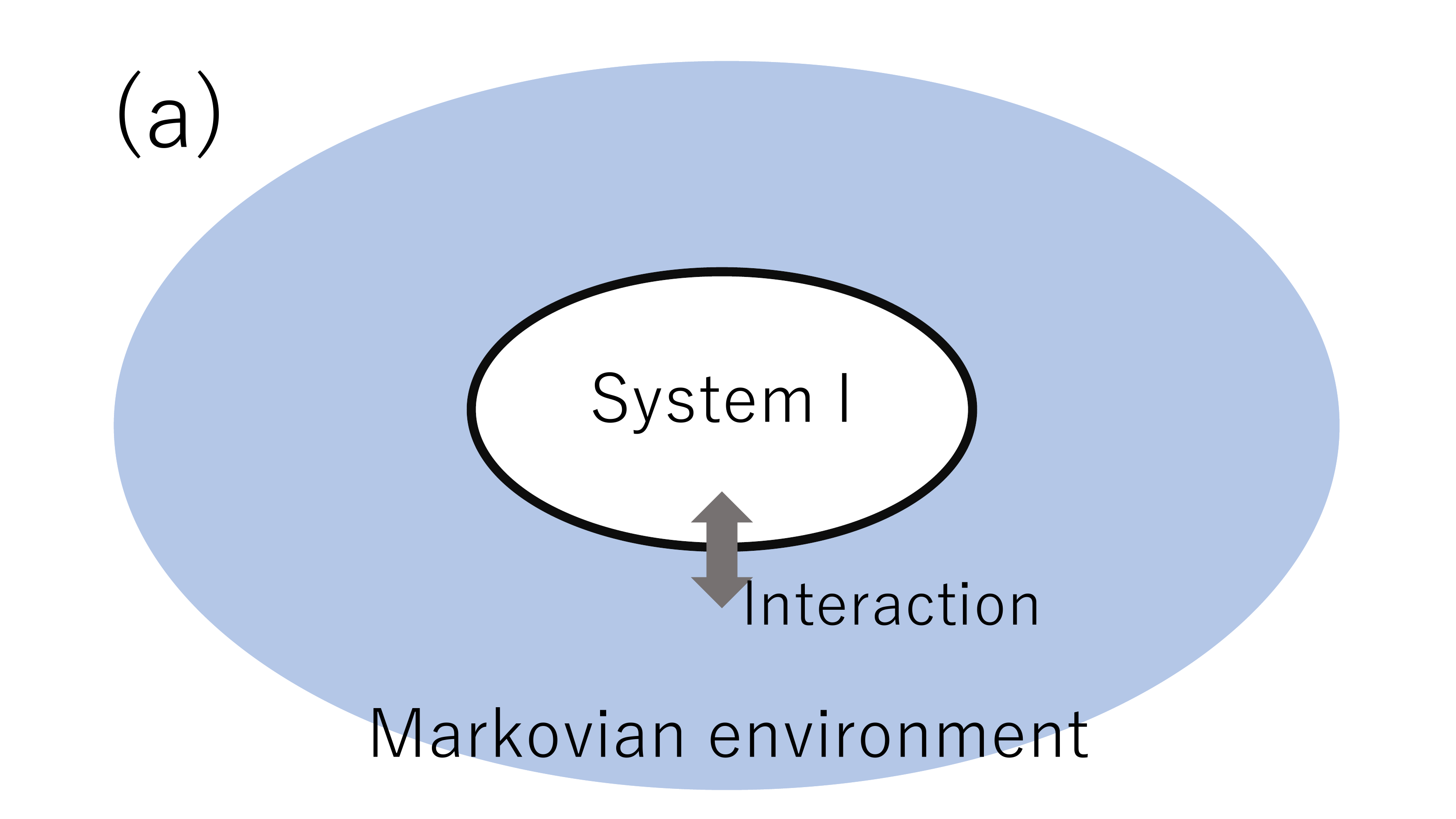}
\end{center}
\end{minipage}
\begin{minipage}[t]{0.48\hsize}
\begin{center}
\includegraphics[width=60mm]{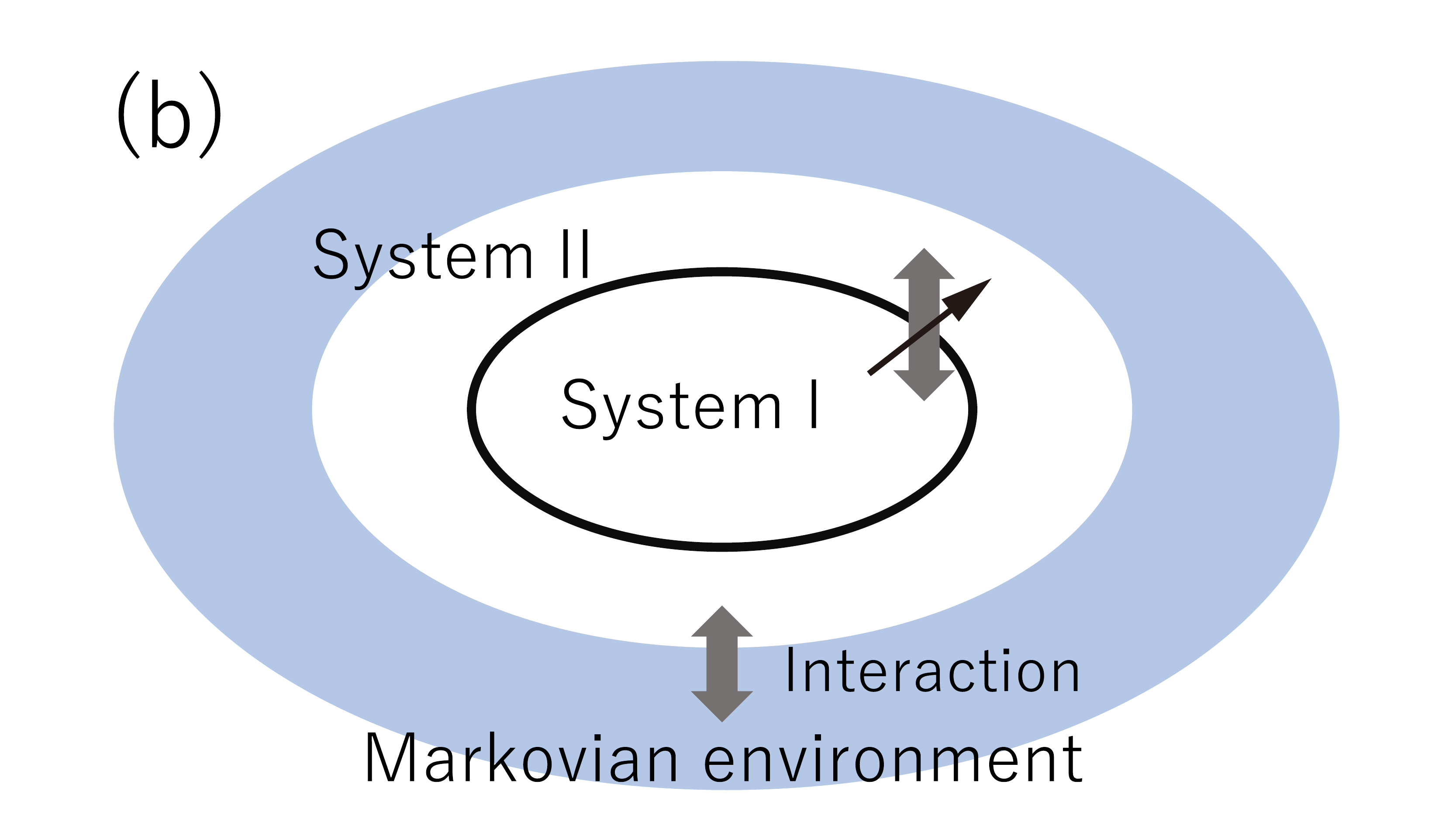}
\end{center}
\end{minipage}
\caption{(a)  System~I interacts with the Markovian environment directly. 
(b) System~I interacts with the Markovian environment indirectly through 
System~II, which results in non-Markovian behaviour in System~I. The arrow symbol
in the coupling of Systems I and II indicates the coupling strength
is effectively variable.
\label{fig:fig1}}
\end{figure}

We propose a theoretical model where non-Markovianity of 
the system dynamics is controlled by adjusting an external field.
The first step is to construct an open system that shows 
non-Markovian dynamics.
This is realised by employing the prescription proposed in
 Ref.~\cite{Ho_2019,kondo2007} as depicted in Fig.~\ref{fig:fig1}.
System~I in Fig.~\ref{fig:fig1}~(a) interacts with the 
environment with a very short-time memory
and shows Markovian relaxation. In Fig.~\ref{fig:fig1}~(b),
System~I is surrounded by System~II, where two systems interact with each other
with a fixed strength. 
While System~II interacts with the Markovian 
environment, System~I interacts with the environment only weakly.  
Hence the main contribution of the relaxation of System~I comes through the
interaction with System~II. 
Relaxation of System~I in this case can be non-Markovian. System~II works 
as a temporal memory and quantum information escaped from System~I is
temporarily stored in System~II before it totally dissipates 
into the environment.
In other words, System~I is in a composite environment (System~II and 
the environment), which has a long-time memory. In the following, 
we consider a case in which System~I is made of one qubit while
System~II is made of $n~( \geq 1)$ identical qubits.

Let us illustrate how to control non-Markovianity before we present detailed 
calculations.
As mentioned before, Systems I and II interact with a fixed strength.
However, the coupling strength can be effectively reduced by applying 
an external field that 
rotates qubits in System~II so that the coupling is partially time-averaged.
In the high-field limit, the coupling strength is totally averaged out and 
System~I suffers only from the Markovian environment.
In this way, it is possible to interpolates between Markovian and non-Markovian 
regimes continuously.

\subsection{Markovian environment}

Let us consider the dynamics of System~I of Fig.~\ref{fig:fig1} 
(a) composed of a single qubit, whose state is given by $\rho$.
See also Ref.~\cite{Ho_2019}.
Dynamics of the qubit as an open quantum system is governed by the 
GKLS master equation \cite{Weiss,Lindblad1976,GKS},
\begin{equation}
\frac{d \rho}{d t}= -i [H, \rho]+{\cal L}[\rho],
\label{eq:dyn}
\end{equation}
where $H$ is the Hamiltonian of System~I and we call ${\cal L}$ 
the Lindbladian, which represents the effect of environment. 
We take $H=0$ here to simplify our analysis.
We use the natural unit $\hbar=1$ throughout this paper.
The Lindbladian for any completely positive semigroup has 
the following form \cite{Lindblad1976,GKS}:
\begin{equation}
{\cal L}[\rho]:=\sum_{i}\gamma_{i}(2 L_{i} \rho 
L^{\dagger}_{i}-\{L_{i}^{\dagger}L_{i},\rho\})
\end{equation}
where $\{ \gamma_{i} \}$ are positive constants.
We consider the case where the environment randomly flip-flops a qubit, 
in which the explicit form of $\mathcal{L}$ is given by
\begin{equation}
{\cal L}[\rho]:=\sum_{\pm}\gamma_\pm 
\left(2\frac{\sigma_{\pm}\rho\sigma_{\mp}}{4}
-\left\{\frac{\sigma_{\mp}\sigma_{\pm}}{4},\rho \right\}\right),
\end{equation}
where $\sigma_{\pm}=(\sigma_{x}\pm i \sigma_{y})/{2}$ and $\sigma_{k}~(k=x,y,z)$ 
are the Pauli matrices~\cite{Weiss}. In this equation, $\gamma_\pm$ represents 
the flip-flop ($| \downarrow \rangle \leftrightarrow |\uparrow \rangle$) rate 
of the qubit  and we assume these rates are symmetric, namely 
$\gamma_+ = \gamma_- :=\gamma_{\rm I}$.

Now GKLS equation is given by
\begin{equation}
\label{eq:aaa}
\frac{d \rho}{d t}=\sum_{\pm}\gamma_{\rm I} 
\left(2\frac{\sigma_{\pm}\rho\sigma_{\mp}}{4}
-\left\{\frac{\sigma_{\mp}\sigma_{\pm}}{4},\rho \right\}\right).
\end{equation}
It is shown that Eq.~(\ref{eq:aaa}) is solved exactly leading 
to exponential relaxation with a characteristic time $2/\gamma_{\rm I}$.

\subsection{Non-Markovian environment: $(1+1)$-qubit case}

We now introduce a theoretical model, in which non-Markovianity can be continuously controlled by an external field.
First, we consider the simplest case where both Systems I and II 
consist of a single qubit, which we call the $(1+1)$-system. 
The System~I qubit has an index 0 while the System~II qubit has an index 1.
The density matrix $\rho^{(1)}$ of the total system is given by
\begin{equation}
\rho^{(1)}=\frac1 2
\begin{pmatrix}
\rho_{11}&\rho_{12}&\rho_{13}&\rho_{14} \\
\rho^{*}_{12}&\rho_{22}&\rho_{23}  &\rho_{24}\\
\rho^{*}_{13}&\rho^{*}_{23}&\rho_{33}&\rho_{34}\\
\rho^{*}_{14}&\rho^{*}_{24}&\rho^{*}_{34}&\rho_{44}\\
\end{pmatrix}
.
\end{equation}
Here the basis vectors are ordered as 
$\{ \ket{00},\ket{01},\ket{10},\ket{11}\}$ 
with $\ket{ab} = \ket{a}_0 \otimes \ket{b}_1, \ a,b \in \{0,1\}$.  
Each qubit in this system is subject to the flip-flop noise independently.
The Lindbladian in this case is given by
\begin{equation}
{\cal L}[\rho^{(1)}]
=\sum_{i={\rm 0,1}} \sum_{\pm}
\gamma_{i}\left(2\frac{\sigma^{(i)}_{\pm}\rho^{(1)}\sigma^{(i)}_{\mp}}{4}
-\left\{\frac{\sigma^{(i)}_{\mp}\sigma^{(i)}_{\pm}}{4}, \rho^{(1)}\right\}
\right)
:=\sum_{i={\rm 0,1}}{\cal L}^{(i)}[\rho^{(1)}],
\end{equation}
where $\sigma^{(i)}_{\mu}$ is the $\mu$-component of the Pauli matrices 
acting non-trivially only on the $i$-th qubit, i.e.,
$\sigma^{(0)}_{\mu}=\sigma_{\mu}\otimes\sigma_{0}$, 
$\sigma^{(1)}_{\mu}=\sigma_{0}\otimes\sigma_{\mu}$ and 
$\sigma^{(i)}_{\pm}={\sigma^{(i)}_{x}\pm i \sigma^{(i)}_{y}}/{2}$ and
$\sigma_{0}$ is the $2\times2$ identity matrix.
Here $\gamma_{i}$ is the flip-flop rate of the $i$-th qubit. 
$\gamma_0 (\gamma_1)$ is also called $\gamma_{\rm I} (\gamma_{\rm II})$ because the $\gamma_{0} (\gamma_1)$ 
is the flip-flop rate of the qubit in System~I (II). 
Suppose the Hamiltonian of the total system is given by
\begin{equation}
H^{(1)} = H_J^{(1)}+ H_{\omega_1}^{(1)}, 
H_J^{(1)} := J\frac{\sigma^{(0)}_{z}\sigma^{(1)}_{z}}{4},
H_{\omega_1}^{(1)} := \omega_{1} \frac{\sigma^{(1)}_{x}}{2}. 
\end{equation}
$H_J^{(1)}$ is a qubit-qubit interaction with a constant strength $J$, while
$H_{\omega_1}^{(1)}$  represents a controllable external 
field $\omega_1$ coupled 
to the $x$-component of the System~II qubit.  

The dynamics of this system is governed by the GKLS master equation,
\begin{align}
\frac{d\rho^{(1)}}{dt}
&=-i[H^{(1)}, \rho^{(1)}]+{\cal L}[\rho^{(1)}]
={\cal D}^{(1)}[\rho^{(1)}]+{\cal L}^{(0)}[\rho^{(1)}],\nonumber\\
{\cal D}^{(1)}[\bullet]
&:=-i\left[ H^{(1)},~\bullet~\right]+{\cal L}^{(1)}[\bullet].
\label{eq:dyn3}
\end{align}
Let us write the density matrix $\rho^{(1)}$ 
in the following form:
\begin{equation}
\rho^{(1)}
=\frac{\sigma^{(0)}_{0}}{2}\cdot\frac{A_{1}^{(1)}+A_{2}^{(1)}}{2}
+\frac{\sigma^{(0)}_{z}}{2}\cdot\frac{A_{1}^{(1)}-A_{2}^{(1)}}{2}
+\frac{\sigma^{(0)}_{+}}{2}\cdot B^{(1)}
+\frac{\sigma^{(0)}_{-}}{2} \cdot (B^{(1)})^{\dagger},
\end{equation}
where
\begin{equation}
A_{1}^{(1)}:=\sigma_{0}\otimes
\begin{pmatrix}
\rho_{11}&\rho_{12}\\
\rho^{*}_{12}&\rho_{22}
\end{pmatrix}
,~
A_{2}^{(1)}:=\sigma_{0}\otimes
\begin{pmatrix}
\rho_{33}&\rho_{34}\\
\rho^{*}_{34}&\rho_{44}
\end{pmatrix}
,~
B^{(1)}:=\sigma_{0}\otimes
\begin{pmatrix}
\rho_{13}&\rho_{14} \\
\rho_{23}  &\rho_{24}
\end{pmatrix}
.
\end{equation}
We easily find that Eq. (\ref{eq:dyn3}) is decomposed 
into the following four equations,
\begin{equation}
\frac{d A_{1}^{(1)}}{d t}=f(A_{1}^{(1)},A_{2}^{(1)}),~
\frac{d A_{2}^{(1)}}{d t}=g(A_{1}^{(1)},A_{2}^{(1)}),~
\frac{d B^{(1)}}{d t}=h(B^{(1)}),~
\frac{d (B^{(1)})^{\dagger}}{d t}=[h(B^{(1)})]^{\dagger}.
\label{eq:form}
\end{equation}
An important observation is that the dynamics of $B^{(1)}$ is decoupled 
from those of $A_{1}^{(1)}$, $A_{2}^{(1)}$ and $(B^{(1)})^{\dagger}$. 

Now we solve Eq.~(\ref{eq:form}) with an appropriate initial condition.
Suppose qubit 0 is polarized along the $x$ axis and qubit 1 is uniformly 
mixed at $t=0$;
\begin{equation}
\rho^{(1)}(0) = \ket{+} \bra{+} \otimes \frac{1}{2} \sigma_0
= \frac{1}{2\cdot2}
\begin{pmatrix}
1&0 & 1&0\\
0 &1&0&1\\
1&0 & 1&0\\
0 &1&0&1
\end{pmatrix}
,
\label{eq:ini1+1}
\end{equation}
where $\ket{+} = \frac{1}{\sqrt{2}}(\ket{0} + \ket{1})$.
This initial condition is rewritten as
\begin{equation}
A_{1}^{(1)}(0)=
A_{2}^{(1)}(0)=B^{(1)}(0)=(B^{(1)})^{\dagger}(0)=\sigma_{0}\otimes
\frac 1 2 \sigma_{0}.
\label{eq:ini}
\end{equation}
Then it turns out that the first two equations in Eq.~(\ref{eq:form}) 
have no dynamics: $f(A_{1}^{(1)},A_{2}^{(1)})=g(A_{1}^{(1)},A_{2}^{(1)})=0$ at any $t$. 
In other words, $A_{1}^{(1)}$ and $A_{2}^{(1)}$ are time-independent 
with this initial condition.
As a result we only need to solve ${d B^{(1)}}/{d t}=h(B^{(1)})$ 
to find the dynamics of the GKLS equation.
To write down the dynamical equation of $B^{(1)}$,
we now evaluate the GKLS equation on 
$\frac{1}{2}\sigma^{(0)}_{+}\cdot B^{(1)}$.
First note that ${\cal L}^{(0)}$ acts only on $\sigma^{(0)}_{+}$ and 
gives just a scalar multiplication:
\begin{equation}
{\cal L}^{(0)}\left[\frac{\sigma^{\rm I}_{+}}{2}\cdot B^{(1)}\right]
=-\Bigl(\frac{\gamma_{\rm I}}{2}\cdot\frac{\sigma^{(0)}_{+}}{2}\Bigr)
\cdot B^{(1)}.
\end{equation}
This implies that $B^{(1)}$ is factorised as 
$B^{(1)}=e^{-\gamma_{\rm I}t/2}\tilde{B}^{(1)}$. 
The dynamics of $\tilde{B}^{(1)}$ following from the GKLS 
equation~(\ref{eq:dyn3}) 
is written as
\begin{equation}
\frac{d}{d t}\Bigl(\frac{\sigma_{+}^{(0)}}{2}\cdot \tilde{B}^{(1)}\Bigr)
={\cal D}^{(1)}\left[ \frac{\sigma_{+}^{(0)}}{2}\cdot \tilde{B}^{(1)}\right].
\label{eq:dyn4}
\end{equation} 
The dynamics of $B^{(1)}$ can be obtained by multiplying $e^{-\gamma_{\rm I} t/2}$ 
to $\tilde{B}^{(1)}$ (or, $B^{(1)}$ with $\gamma_{\rm I} = 0$). 
Therefore, we will consider the case when $\gamma_{\rm I} = 0$ hereafter. 

$\tilde{B}^{(1)}$ can be expanded as 
$\tilde{B}^{(1)}=\frac{1}{2}\sum_{k=0,x,y,z}b_{k} \sigma^{(1)}_{k}$, where
\begin{equation}
b_{0}=\rho_{13}+\rho_{24},~
b_{x}=\rho_{14}+\rho_{23},~
b_{y}=i(\rho_{14}-\rho_{23}),~
b_{z}=\rho_{13}-\rho_{24}.
\nonumber
\end{equation}
We now evaluate the right-hand side of Eq. (\ref{eq:dyn4}).
The action on each basis $\frac{1}{2}\sigma^{(0)}_{+}\cdot(\sigma^{(1)}_{k}/2)$ 
of $\frac{1}{2}\sigma^{(0)}_{+}\cdot\tilde{B}^{(1)}$ is given as
\begin{align}
{\cal D}^{(1)}\left[ \frac{\sigma_{+}^{(0)}}{2}\cdot 
\frac{\sigma^{(1)}_{0}}{2} \right]
&=- i \frac{J}{2}\Bigl( \frac{\sigma_{+}^{(0)}}{2}\cdot
\frac{\sigma^{(1)}_{z}}{2} \Bigr),\nonumber\\
{\cal D}^{(1)}\left[ \frac{\sigma_{+}^{(0)}}{2}\cdot 
\frac{\sigma^{(1)}_{x}}{2} \right]
&=-\frac{\gamma_{\rm II}}{2}\Bigl( \frac{\sigma_{+}^{(0)}}{2}\cdot
\frac{\sigma^{(1)}_{x}}{2} \Bigr),\nonumber\\
{\cal D}^{(1)}\left[ \frac{\sigma_{+}^{(0)}}{2}\cdot 
\frac{\sigma^{(1)}_{y}}{2} \right]
&= \omega_{1}\Bigl( \frac{\sigma_{+}^{(0)}}{2}\cdot
\frac{\sigma^{(1)}_{z}}{2} \Bigr)-\frac{\gamma_{\rm II}}{2}
\Bigl( \frac{\sigma_{+}^{(0)}}{2}\cdot\frac{\sigma^{(1)}_{y}}{2} \Bigr),\nonumber\\
{\cal D}^{(1)}\left[  \frac{\sigma_{+}^{(0)}}{2}\cdot 
\frac{\sigma^{(1)}_{z}}{2} \right]
&=- i \frac{J}{2}\Bigl( \frac{\sigma_{+}^{(0)}}{2}\cdot\frac{\sigma^{(1)}_{0}}{2} 
\Bigr)-\gamma_{\rm II}\Bigl( \frac{\sigma_{+}^{(0)}}{2}\cdot
\frac{\sigma^{(1)}_{z}}{2} \Bigr)- \omega_{1}
\Bigl( \frac{\sigma_{+}^{(0)}}{2}\cdot\frac{\sigma^{(1)}_{y}}{2} \Bigr).
\end{align}
We summarise the action of ${\cal D}^{(1)}$ on $\frac{1}{2}\sigma_{+}^{(0)} \cdot \tilde{B}^{(1)}$ as
\begin{equation}
{\cal D}^{(1)} \left[ \frac{\sigma_{+}^{(0)}}{2}\cdot \tilde{B}^{(1)} \right] 
= \frac{\sigma_{+}^{(0)}}{2} \cdot \sum_{k,m=0,x,y,z}b_{k}
({\bold M}_{0})_{km}\frac{\sigma^{(i)}_{m}}{2},
\end{equation}
where
\begin{equation}
{\bold M_{0}}:=\frac{1}{2}
\begin{matrix}
0~~~~~x~~~~~~y~~~~~~z\\
\begin{pmatrix}
0&0&0&- i J\\
0&-\gamma_{\rm II}&0&0\\
0&0&- \gamma_{\rm II}&2 \omega_{1}\\
- i J&0&-2 \omega_{1}&-2\gamma_{\rm II}
\end{pmatrix}
.
\end{matrix}
\end{equation}
Comparing the coefficients of each basis 
$\frac{1}{2}\sigma^{(0)}_{+}\cdot(\sigma^{(1)}_{k}/2)$ 
in the left-hand side and the right-hand side of Eq. (\ref{eq:dyn4}), 
we obtain the following differential equations for $b_{k}$:
\begin{equation}
\frac{d}{d t}
\begin{pmatrix}
b_{0}\\
b_{x}\\
b_{y}\\
b_{z}
\end{pmatrix}
=
{\bold M_{0}}^{\rm T}
\begin{pmatrix}
b_{0}\\
b_{x}\\
b_{y}\\
b_{z}
\end{pmatrix}
.
\end{equation}

Note that  the dynamics of $b_{x}$ is totally decoupled from the 
other variables.
Hereafter, we ignore the dynamics of $b_{x}$ by employing the initial 
condition (\ref{eq:ini}),
that is, $b_0(0)=1$ and $b_x(0)=b_y(0)=b_z(0)=0$.
The remaining equations are concisely written in the following matrix form:
\begin{equation}
\frac{d}{d t}
\begin{pmatrix}
b_{0}\\
b_{y}\\
b_{z}
\end{pmatrix}
={\bold M}^{\rm T}
\begin{pmatrix}
b_{0}\\
b_{y}\\
b_{z}
\end{pmatrix}
,
\label{eq:key}
\end{equation}
where
\begin{equation}
{\bold M}:=\frac{1}{2}
\begin{pmatrix}
0&0&- i J\\
0&-\gamma_{\rm II}&2 \omega_{1}\\
- i J&-2 \omega_{1}&-2 \gamma_{\rm II}
\end{pmatrix}
.
\end{equation}
This equation is analytically solvable since $\mathbf{M}$ is constant 
and its eigenvalues and eigenvectors are easily found (See Appendix A).

Let us evaluate the reduced density matrix of System~I, $\rho_{\rm I}^{(1)}$ 
by tracing out System~II with the initial condition (\ref{eq:ini}).
After straightforward calculation, we obtain
\begin{equation}
\rho^{(1)}_{{\mathrm I}}
= {\rm Tr_{II}}(\rho^{(1)})
=\frac{1}{2}
\begin{pmatrix}
1&e^{-\gamma_{\rm I}t/2}b_{0}(t)\\
e^{-\gamma_{\rm I}t/2}b_{0}(t)&1
\end{pmatrix}
.
\end{equation}
The explicit form of $b_0(t)$ is given in Appendix A, where we also show that $b_{0}(t)$ is real.
Note that  $\rho^{(1)}_{{\mathrm I}}(0) = \ket{+} \bra{+}$ and 
$\rho^{(1)}_{{\mathrm I}}(\infty)= \sigma_0/2$.

\subsection{Non-Markovian environment: $(1+n)$-qubit case}

The above analysis is readily generalised to the case where System~II 
consists of $n$ identical qubits.
We call this system the $(1+n)$-system \cite{Ho_2019}.
We consider a system in which
the System~I qubit interacts with all System~II qubits 
with equal coupling strength $J$
while the qubits in System~II do not interact among themselves.   
Moreover, there is an external field $\omega_1$ that couples equally with all 
the System~II qubits.
The Hamiltonian of this system is then given by
\begin{align}
H=\sum^{n}_{i=1}\left( H^{(i)}_{J}+H^{(i)}_{\omega_{1}}\right),
\ H^{(i)}_{J}:=J\frac{\sigma^{(0)}_{z}\sigma^{(i)}_{z}}{4},\ 
H^{(i)}_{\omega_{1}}:=\omega_{1} \frac{\sigma^{(i)}_{x}}{2},
\label{eq:hami2}
\end{align}
where 
$\sigma_{\mu}^{(i)} = \sigma_0 \otimes \ldots \otimes \sigma_0 
\otimes \sigma_{\mu} \otimes \sigma_0 \otimes  \ldots \otimes \sigma_0$ nontrivially acts only on the $i$-th qubit.
Here we assign an index 0 to the System~I qubit
while indices 1 to $n$ to the System~II qubits.
The basis vectors are ordered as
\begin{align}
\{ \ket{00 \ldots 00},\ket{00 \ldots 01}, \ket{00\ldots 11}, \ldots ,
\ket{11 \ldots 10},
\ket{11 \ldots 11}\},
\end{align}
where $\ket{ab \ldots cd}= \ket{a}_0 \otimes \ket{b}_1\otimes  \ldots \otimes
\ket{c}_{n-1} \otimes \ket{d}_n$. 

The Lindbladian which represents the flip-flop noise that acts on all qubits 
independently is
\begin{equation}
{\cal L}[\rho]=\sum^{n}_{i=0} \sum_{\pm}\gamma_{i}
\left(2\frac{\sigma^{(i)}_{\pm}\rho\sigma^{(i)}_{\mp}}{4}
-\left\{\frac{\sigma^{(i)}_{\mp}\sigma^{(i)}_{\pm}}{4}, \rho\right\}\right)
:=\sum^{n}_{i=0}{\cal L}^{(i)}[\rho].
\label{eq:lind2}
\end{equation}
We assume from now on that the strength $\gamma_{i}$ for all the qubits in System~II 
are identical:
$\gamma_1=\gamma_{2}=\ldots=\gamma_{n}=:\gamma_{\rm II}$.

The dynamics of the density matrix $\rho^{(n)}$ of Systems~I and II 
is described by
\begin{align}
\frac{d\rho^{(n)}}{dt}&=-i[H, \rho^{(n)}]+{\cal L}[\rho^{(n)}]
=\sum^{n}_{i=1}{\cal D}^{(i)}[\rho^{(n)}]+{\cal L}^{(0)}[\rho^{(n)}],\nonumber\\
{\cal D}^{(i)}[\bullet]&:=-i\left[ (H^{(i)}_{J}
+H^{(i)}_{\omega_{1}}),~\bullet~\right]+{\cal L}^{(i)}[\bullet].
\label{eq:dyn2}
\end{align}

Let us write $\rho^{(n)}$ in the same  form as the $(1+1)$-case, 
\begin{equation}
\rho^{(n)}
=\frac{\sigma^{(0)}_{0}}{2}\cdot\frac{A_{1}^{(n)}+A_{2}^{(n)}}{2}
+\frac{\sigma^{(0)}_{z}}{2}\cdot\frac{A_{1}^{(n)}-A_{2}^{(n)}}{2}
+\frac{\sigma^{(0)}_{+}}{2}\cdot B^{(n)}
+\frac{\sigma^{(0)}_{-}}{2} \cdot (B^{(n)})^{\dagger}.
\end{equation}
$A_1^{(n)}$, $A_2^{(n)}$ and $B^{(n)}$ respectively have matrix forms 
$\sigma_{0}\otimes {A_{1}'}^{(n)}$, $\sigma_{0}\otimes {A'_{2}}^{(n)}$ 
and $\sigma_{0}\otimes {B'}^{(n)}$ where ${A'_1}^{(n)}$, ${A'_2}^{(n)}$ 
and ${B'}^{(n)}$ are $2^{n}\times2^{n}$ matrices.
Equivalently, $\rho^{(n)}$ can be represented by the following block matrix form:
\begin{equation}
\rho^{(n)}=
\begin{pmatrix}
{A'_1}^{(n)} &{B'}^{(n)}\\
({B'}^{(n)})^{\dagger}& {A'_{2}}^{(n)}
\end{pmatrix}
.
\end{equation}
We can find that the dynamics of $B^{(n)}$ is decoupled from those of 
$A_{1}^{(n)}$, $A_{2}^{(n)}$ and $(B^{(n)})^{\dagger}$ as in the (1+1)-case.
We are interested in the initial state
\begin{equation}
\rho^{(n)}(0) = \ket{+} \bra{+} \otimes (\frac{1}{2} \sigma_0)^{\otimes n}
= \frac{1}{2^{n+1}}
\begin{pmatrix}
\sigma^{\otimes n}_{0}&\sigma^{\otimes n}_{0}\\
\sigma^{\otimes n}_{0}&\sigma^{\otimes n}_{0}
\end{pmatrix}
,
\label{eq:ini1+n}
\end{equation}
which is a generalisation of Eq.~(\ref{eq:ini1+1}) 
for the $(1+1)$-system to the $(1+n)$-system. 
This initial condition in terms of $A_1^{(n)}, A_2^{(n)}, B^{(n)}$ 
and $(B^{(n)})^{\dagger}$ is
\begin{equation}
A_{1}^{(n)}(0)=
A_{2}^{(n)}(0)= B^{(n)}(0) =(B^{(n)})^{\dagger}(0)=
\sigma_{0}\otimes\frac{1}{2^{n}}\sigma^{\otimes n}_{0}\qquad (t\geq 0)
\end{equation}
Solutions of $A_1^{(n)}$ and $A_2^{(n)}$ are
time independent with this initial condition.

Since the action of ${\cal L}^{(0)}$ gives just a scalar multiplication 
as mentioned previously, we find that 
the GKLS equation~(\ref{eq:dyn2}) can be rewritten as
\begin{equation}
\frac{d}{d t}\Bigl(\frac{\sigma_{+}^{(0)}}{2}\cdot\tilde{B}^{(n)}\Bigr)
=\sum^{n}_{i=1}{\cal D}^{(i)}\left[ \frac{\sigma_{+}^{(0)}}{2}\cdot 
\tilde{B}^{(n)}\right],
\label{eq:dyn5}
\end{equation}
where $\tilde{B}^{(n)}:=e^{-\gamma_{\rm I}t/2}B^{(n)}$.
We write
\begin{equation}
\tilde{B}^{(n)}= 
\prod^{n}_{i=1}\varsigma^{(i)}
\label{eq:bpart}
\end{equation}
where 
\begin{equation}
\varsigma^{(i)}=\frac{1}{2} \sum_{k=0,x,y,z} b^{(i)}_{k} \sigma_{k}^{(i)}
\label{eq:expression}
\end{equation}
with $b^{(i)}_{k} \in \mathbb{C}$. Our initial condition gives 
$b^{(i)}_{0}(0)=1, b^{(i)}_{x}(0)=b^{(i)}_{y}(0)=b^{(i)}_{z}(0)=0$ where $1\leq i \leq n$. 
It turns out that $b^{(i)}_{x}$ decouples from the dynamics of the other 
$b^{(i)}_{k}$'s and we can set $b^{(i)}_{x}(t)= 0$ from the given initial condition.
It follows from Eq.~(\ref{eq:bpart}) that the density matrix
$\rho^{(n)}$ correctly 
reflects the symmetry under arbitrary permutation of $n$ qubits in System~II
and that there are no correlations among them. 

We then calculate the action of ${\cal D}^{(i)}$.
Since ${\cal D}^{(i)}$ acts only on the $0$-th and $i$-th qubits,
it sufficies to consider the term 
$\frac{1}{2}\sigma_{+}^{(0)}\cdot \varsigma^{(i)}$ only.
The action of ${\cal D}^{(i)}$ on 
$\frac{1}{2}\sigma_{+}^{(0)}\cdot ({\sigma^{(i)}_{\mu}}/{2})$ is given as
\begin{align}
{\cal D}^{(i)}\left[ \frac{\sigma_{+}^{(0)}}{2}\cdot 
\frac{\sigma^{(i)}_{0}}{2} \right]&=- i \frac{J}{2}
\Bigl( \frac{\sigma_{+}^{(0)}}{2}\cdot
\frac{\sigma^{(i)}_{z}}{2} \Bigr),\nonumber\\
{\cal D}^{(i)}\left[ \frac{\sigma_{+}^{(0)}}{2}\cdot 
\frac{\sigma^{(i)}_{y}}{2} \right]&= \omega_{1}
\Bigl( \frac{\sigma_{+}^{(0)}}{2}\cdot\frac{\sigma^{(i)}_{z}}{2} \Bigr)
-\frac{\gamma_{\rm II}}{2}\Bigl( \frac{\sigma_{+}^{(0)}}{2}\cdot
\frac{\sigma^{(i)}_{y}}{2} \Bigr),\nonumber\\
{\cal D}^{(i)}\left[  \frac{\sigma_{+}^{(0)}}{2}\cdot 
\frac{\sigma^{(i)}_{z}}{2} \right]&=- i \frac{J}{2}
\Bigl( \frac{\sigma_{+}^{(0)}}{2}\cdot\frac{\sigma^{(i)}_{0}}{2} \Bigr)
-\gamma_{\rm II} \Bigl( \frac{\sigma_{+}^{(0)}}{2}\cdot
\frac{\sigma^{(i)}_{z}}{2} \Bigr)- \omega_{1}
\Bigl( \frac{\sigma_{+}^{(0)}}{2}\cdot\frac{\sigma^{(i)}_{y}}{2} \Bigr).
\end{align}
We summarise the action of ${\cal D}^{(i)}$ on 
$ \frac{\sigma_{+}^{(0)}}{2}\cdot \varsigma^{(i)}$ as
\begin{equation}
{\cal D}^{(i)}\left[ \frac{\sigma_{+}^{(0)}}{2}\cdot 
\varsigma^{(i)} \right]= \frac{\sigma_{+}^{(0)}}{2}\cdot
\sum_{k,m=0,y,z}b^{(i)}_{k}({\bold M})_{km}\frac{\sigma^{(i)}_{m}}{2},
\end{equation}
where
\begin{equation}
{\bold M}=\frac{1}{2}
\begin{matrix}
0~~~~~y~~~~~z\\
\begin{pmatrix}
0&0&- i J\\
0&- \gamma_{\rm II}&2 \omega_{1}\\
- i J&-2 \omega_{1}&-2\gamma_{\rm II}
\end{pmatrix}
\end{matrix}
,
\end{equation}
which is the same ${\bold M}$ introduced in the $(1+1)$-system. 
The dynamics of $\tilde{B}^{(n)}$ following from 
Eq.~(\ref{eq:dyn5}) is written as
\begin{align}
\frac{d}{d t}\Bigl(\frac{\sigma_{+}^{(0)}}{2}\cdot&\prod^{n}_{i=1}
\varsigma^{(i)}\Bigr)
=\sum^{n}_{i=1}{\cal D}^{(i)}\left[ \frac{\sigma_{+}^{(0)}}{2}\cdot 
\Bigl(\prod^{n}_{i=1}\varsigma^{(i)}\Bigr)\right]\nonumber\\
=&\frac{\sigma_{+}^{(0)}}{2}\cdot\left[ \Bigl(\sum_{k,m=0,y,z}b^{(1)}_{k}
({\bold M})_{km}\frac{\sigma^{(1)}_{m}}{2}\Bigr)\cdot\varsigma^{(2)}
\cdot \ldots\cdot\varsigma^{(i)}\cdot \ldots\cdot\varsigma^{(n)}\right]
+\ldots\nonumber\\
&+\ldots+ \frac{\sigma_{+}^{(0)}}{2}\cdot\left[\varsigma^{(1)}\cdot 
\ldots\cdot\Bigl(\sum_{k,m=0,y,z}b^{(i)}_{k}({\bold M})_{km}
\frac{\sigma^{(i)}_{m}}{2}\Bigr)\cdot \ldots\cdot\varsigma^{(n)}\right]
\nonumber\\
&+\ldots+ \frac{\sigma_{+}^{(0)}}{2}\cdot\left[ \varsigma^{(1)}\cdot 
\ldots\cdot\varsigma^{(i)}\cdot \ldots\cdot\Bigl(\sum_{k,m=0,y,z}b^{(n)}_{k}
({\bold M})_{km}\frac{\sigma^{(n)}_{m}}{2}\Bigr)\right].
\end{align}
Comparing the coefficients of each basis in the left-hand side and 
the right-hand side, we obtain differential equations 
for each coefficient $b^{(1)}_{k}\ldots b^{(i)}_{k}\ldots b^{(n)}_{k}$ as
\begin{align}
\frac{d}{d t} (b^{(1)}_{k}\ldots b^{(i)}_{k}\ldots b^{(n)}_{k})
=&\Bigl(\sum_{m=0,y,z}({\bold M}^{\rm T})_{km}b^{(1)}_{m}\Bigr)
b^{(2)}_{k}\ldots b^{(i)}_{k}\ldots b^{(n)}_{k}+\ldots\nonumber\\
+&b^{(1)}_{k}\ldots b^{(i-1)}_{k}\Bigl(\sum_{m=0,y,z}
({\bold M}^{\rm T})_{km}b^{(i)}_{m}\Bigr)b^{(i+1)}_{k}\ldots 
b^{(n)}_{k}+\ldots\nonumber\\
+&b^{(1)}_{k}\ldots b^{(i)}_{k}\ldots b^{(n-1)}_{k}\Bigl(\sum_{m=0,y,z}
({\bold M}^{\rm T})_{km}b^{(n)}_{m}\Bigr).
\label{eq:n}
\end{align}
We rewrite this equation as
\begin{align}
&\Bigl(\frac{d  b^{(1)}_{k}}{d t}-\sum_{m=0,y,z}
({\bold M}^{\rm T})_{km} b^{(1)}_{m}\Bigr) b^{(2)}_{k}\ldots b^{(i)}_{k}
\ldots b^{(n)}_{k}+\ldots\nonumber\\
+& b^{(1)}_{k}\ldots b^{(i-1)}_{k}\Bigl(\frac{d  b^{(i)}_{k}}{d t}-\sum_{m=0,y,z}
({\bold M}^{\rm T})_{km} b^{(i)}_{m}\Bigr) b^{(i+1)}_{k}\ldots b^{(n)}_{k}
+\ldots\nonumber\\
+& b^{(1)}_{k}\ldots b^{(i)}_{k}\ldots b^{(n-1)}_{k}\Bigl(\frac{d  b^{(n)}_{k}}{d t}
-\sum_{m=0,y,z}({\bold M}^{\rm T})_{km} b^{(n)}_{m}\Bigr)=0.
\end{align}
We obtain the differential equations 
\begin{equation}
\frac{d}{d t}
\begin{pmatrix}
 b^{(i)}_{0}\\
 b^{(i)}_{y}\\
 b^{(i)}_{z}
\end{pmatrix}
={\bold M}^{\rm T}
\begin{pmatrix}
 b^{(i)}_{0}\\
 b^{(i)}_{y}\\
 b^{(i)}_{z}
\end{pmatrix}
, \qquad 1 \leq i \leq n. 
\label{eq:ncasedynamics}
\end{equation}
Note that these differential equations are the same as Eq.~(\ref{eq:key}) 
in the $(1+1)$-system. Moreover, the initial conditions are the same for 
all $i$ and thus the dynamics is solvable for any $n$ 
by employing $b_0(t)$ obtained for the (1+1)-system. This solution 
is reasonable because the qubits in System~II are identical. 

Let us evaluate the reduced density matrix $\rho_{\rm I}^{(n)}$ of System~I, 
by tracing out System~II.
Note that $\tr(\varsigma^{(i)})=b_{0}$ since Pauli matrices are traceless.
We obtain
\begin{align}
\rho^{(n)}_{\mathrm I}(t):=\tr_{\rm II}\, \rho^{(n)}
=&\frac{\sigma^{(0)}_{0}}{2}\cdot \tr\left(\prod^{n}_{i=1}
\frac{\sigma^{(i)}_{0}}{2}\right)+e^{-\gamma_{\rm I}t/2}
\left[\frac{\sigma^{(0)}_{+}}{2}\cdot\tr\left(\prod^{n}_{i=1}
\varsigma^{(i)}\right)+
\mbox{h.c.}\right]\nonumber\\
=&\frac{\sigma^{(0)}_{0}}{2}+e^{-\gamma_{\rm I}t/2}
\left(\frac{\sigma^{(0)}_{+}}{2}b_{0}^n+\mbox{h.c.}\right)
\nonumber\\
=& \frac{1}{2}
\begin{pmatrix}
1&e^{-\gamma_{\rm I}t/2}\left(b_{0} (t)\right)^{n}\\
e^{-\gamma_{\rm I}t/2}\left(b_{0}(t)\right)^{n}&1
\end{pmatrix}.
\label{eq:reddens}
\end{align}
We find that the effect of the direct coupling of the Markovian environment 
with System~I, shown as the factor of $e^{-\gamma_{\rm I} t/2}$, is well 
separated from those through System~II, which is the origin of the 
non-Markovian dynamics of System~I. 
We introduce 
\begin{align}
 \beta_{n}(t):=e^{-\gamma_{\rm I}t/2}(b_{0}(t))^{n} 
\end{align}
for later convenience.

One might think that our model is a trivial extension of one introduced 
in Ref.~\cite{Ho_2019}
since the only difference is the existence of the external field $\omega_1$.
However, $H_{\omega_1}^{(i)}$ does not commute with $H_J^{(i)}$, which
makes our solution highly non-trivial compared to that obtained in  
Ref.~\cite{Ho_2019}.
Moreover, the dynamics of our model has three degrees of freedom,  
while that of Ref.~\cite{Ho_2019} has only two degrees of freedom. As a result,
this model shows drastically different behaviours from the previous one.
It is possible to control non-Markovianity of the dynamics continuously 
by changing the external field strength $\omega_1$ as will be shown 
in Sec.~\ref{sec:3}.

\subsection{Non-Markovianity measure}

We will discuss control of non-Markovianity of dynamics 
by manipulating an external field in Sec.~\ref{sec:3}. 
To this end, let us first introduce a measure ${\cal N}$ to quantify non-Markovianity 
of dynamics.
We employ the measure proposed in \cite{PhysRevLett.103.210401},
which is based on the concept of {\it information backflow} 
from the environment.
This measure is described in terms of the trace distance 
$D[\rho(t),\rho'(t)]=\tr|\rho(t)-\rho'(t)|/2$ 
between two states $\rho$ and $\rho'$ of a system of interest.
Note that the environmental freedoms are traced out here.
${\cal N}$ introduced in \cite{PhysRevLett.103.210401} 
is defined as 
\begin{equation}
{\cal N} :=\max_{\rho(0),\rho'(0)}\int_{\Omega_{+}} 
\frac{d D[\rho(t),\rho'(t)]}{dt}dt,
\label{eq:original}
\end{equation}
where $\Omega_{+}:= \{t \in [0, \infty)| dD[\rho(t),\rho'(t)]/dt \geq 0 \}$
is a disjoint union of many intervals in general.

In this study, let us restrict the maximisation in ${\cal N}$ 
with respect to the initial System~I states written as 
\begin{equation}
\rho_{\rm I}^{(n)}(t=0,\theta)
=\frac12
\begin{pmatrix}
1&e^{i \theta}\\
e^{-i \theta}&1
\end{pmatrix},~\theta \in {\mathbb R}.
\label{eq:ini2}
\end{equation}
Note that $\rho_{\rm I}^{(n)}(t=0,\theta)$ corresponds the following 
initial state of System~I and II since the initial state of System~II
is fixed to $\displaystyle \prod^{n}_{i=1}\frac{\sigma^{(i)}_{0}}{2}$. 
\begin{equation}
\rho^{(n)}(t=0,\theta)
=\frac12
\begin{pmatrix}
1&e^{i \theta}\\
e^{-i \theta}&1
\end{pmatrix}
\otimes \Bigl(\prod^{n}_{i=1}\frac{\sigma^{(i)}_{0}}{2} \Bigr),
\label{eq:form1}
\end{equation}
The dynamics of the reduced density matrix of System~I starting 
from the above initial state can be written as
\begin{equation}
\rho_{\rm I}^{(n)}(t, \theta)
=\frac12
\begin{pmatrix}
1&e^{i \theta} \beta_n(t)\\
e^{-i \theta} \beta_n (t)& 1 
\end{pmatrix}
.
\label{eq:theta}
\end{equation}
Thus, we calculate the trace distance between any two states initially written 
as Eq.~(\ref{eq:ini2}):
\begin{align}
D[\rho_{\rm I}^{(n)}(t,\theta_1),\rho_{\rm I}^{(n)}(t,\theta_2)]
&=\Big|\beta_{n}(t)
\sin\Bigl(\frac{\theta_{1}-\theta_{2}}{2}\Bigr)\Big|.
\end{align}
A pair of pure states in System~I 
with antipodal initial Bloch vectors, $\rho_{\rm I}^{(n)}(0,\theta)$ and  
$\rho_{\rm I}^{(n)}(0,\theta+\pi)$, 
gives the maximum value of the integrand 
$d D[\rho_{\rm I}^{(n)}(t, \theta),\rho_{\rm I}^{(n)}(t, \theta+\pi)]/d t$ 
at any $t>0$. Thus, ${\cal N}$ is rewritten as
\begin{equation}
{\cal N}=\int_{\Omega_{+}}dt \frac{d D[\rho_{\rm I}^{(n)}(t,\theta),
\rho_{\rm I}^{(n)}(t,\theta+\pi)]}{d t}
=\int_{\Omega_{+}}dt \frac{d |\beta_{n}(t)|}{d t}.
\label{eq:nonmarkov2}
\end{equation}

In Sec.~\ref{sec:3}, we evaluate ${\cal N}$ and compare them 
with those obtained by NMR experiment.

\section{non-Markovianity Control: Experiment}
\label{sec:3}
\subsection{Experimental Setup and Hamiltonian}

In Sec.~\ref{sec:2}, we conducted theoretical analysis of 
a fictitious system that is made of
one-qubit System~I, identical $n$-qubit System~II and Markovian environment. 
In this section, we map
this model to a molecular system that can be realised in liquid-state NMR. 
We briefly introduce this system to make our work self-contained. 
See Ref.~\cite{Ho_2019} for further details. 

\begin{figure}[h]
\begin{center}
\includegraphics[width=100mm]{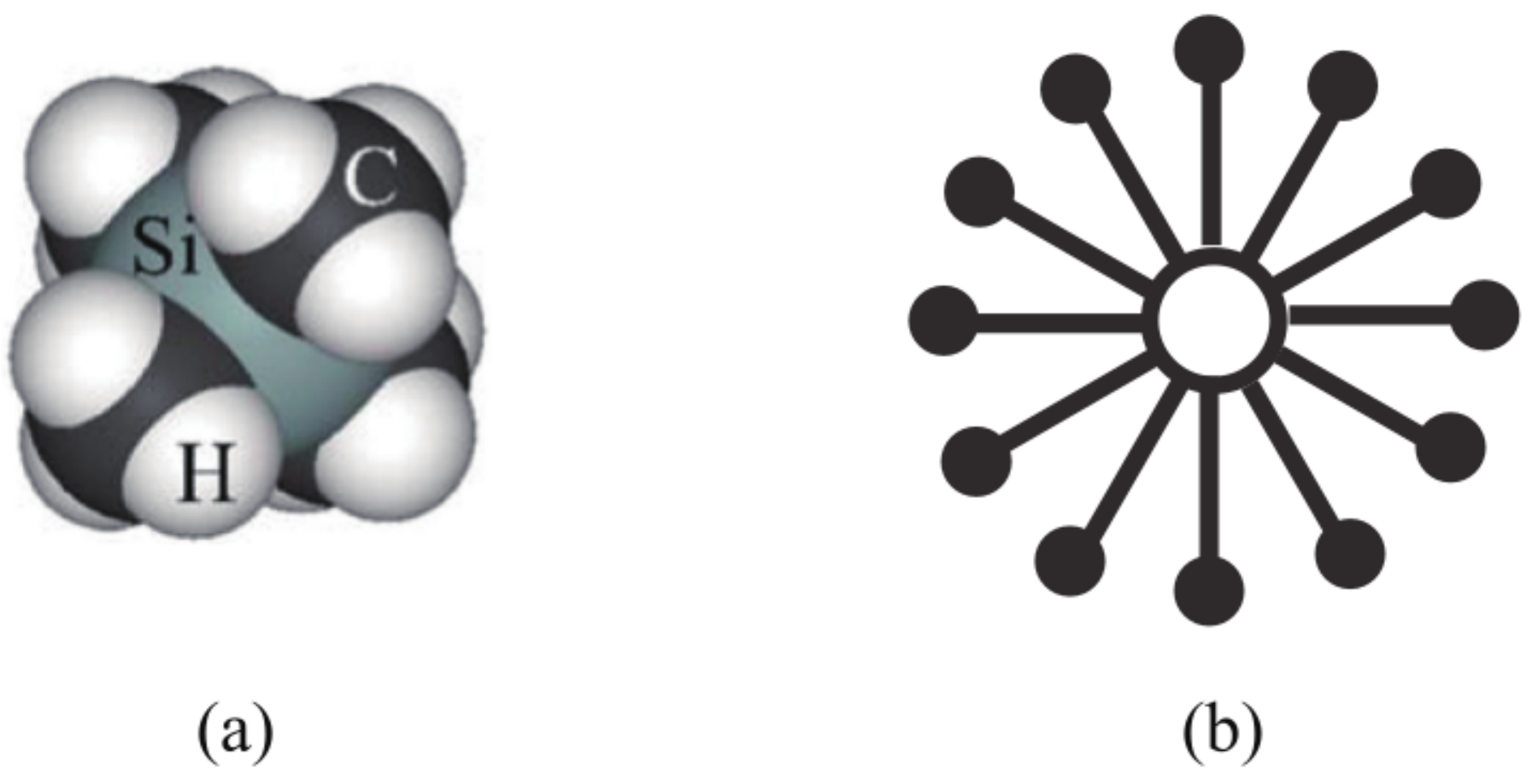}
\end{center}
\caption{(a) Tetramethylsilane (TMS) molecule. 
(b) Schematic picture of the star-topology spin network of TMS. 
The open circle is System~I ($^{29}$Si nucleus) while the
filled circles form System~II (H nuclei). 
\label{fig:starshape}}
\end{figure}

In NMR, a spin-1/2 nucleus is identified with a qubit. Under a strong 
magnetic field, the nucleus
has a well-defined spin-up (spin-down) state that corresponds 
to $\ket{0}$ ($\ket{1}$) qubit state.
We take a star-topology molecule for Systems I and II, in which System~I 
is the central nucleus while System~II is formed by the surrounding nuclei, 
see Fig.~\ref{fig:starshape}.
We consider a molecule in which the nucleus of System~I and nuclei of 
System~II belong to different nuclear species while all nuclei 
in System~II are identical.
Because of the symmetry of System II, System~I nucleus interact 
with each nucleus
of System~II with equal strength $J$. Interactions among nuclei of System~II
effectively vanish because of symmetry and motional narrowing \cite{Levitt2008}.
In addition, an external RF (radio frequency) magnetic field 
is applied on the molecule. 
If the RF frequency is equal to the Larmor frequency of the spins in System~II,
it acts as a static external field for the spins in System~II, 
while it has no effect on the spin in System~I
in the rotating frame of respective nuclei.
As a result, the Hamiltonian of System~I and II is approximated by
\begin{align}
H&=J\sum^{n}_{i=1}\frac{\sigma^{(0)}_{z}\sigma^{(i)}_{z}}{4}
+\omega_{1} \sum^{n}_{i=1}\frac{\sigma^{(i)}_{x}}{2}.
\label{eq:hami3}
\end{align}
which reproduces Eq.~(\ref{eq:hami2}). Here $J$ is the common coupling 
strength between the System~I spin 
and the System~II spins while $\omega_1$ is a measure of the RF magnetic  
field amplitude.
 
We employed Tetramethylsilane (TMS, ${\rm C}_{4}{\rm H}_{12}{\rm Si}$) 
as such a molecule in our experiment.
A TMS molecule is a star-topology molecule that corresponds 
to the $(1,12)$~system (Fig. \ref{fig:starshape}). 
The central nucleus of $^{29}$Si acts as System~I while surrounding 
12 hydrogen nuclei form System~II. 
Molecules are solved in acetone-d6 that is isotropic~\cite{Levitt2008}.
The spin flip-flop rates $\gamma_{\rm I}$ and $\gamma_{\rm II}$ 
can be controlled by adding some magnetic impurities into the sample 
solution, see Ref.~\cite{Ho_2019,Kondo_2016,Iwakura2017,kondo2018}.
Although we did not intentionally add the magnetic impurities 
into the solvent in our experiment, oxygen molecules 
in the solvent act as the magnetic impurities.
In NMR experiments, we observe Free Induction Decay signals 
(FID's hereinafter) that represent the relaxation of the expectation 
value of $\sigma_{x}$ and $\sigma_{y}$ (strictly speaking, 
it is an ensemble average over many TMS molecules).
In our model, this relaxation is described with 
$\beta_{n}(t)$. 
To compare the theoretical and experimental results, we first 
measured $\gamma_{\rm I}$ by fitting the decoupling (Markovian) 
limit of the experimental data with a function $e^{-\gamma_{\rm I} t/2}$.
We independently measured $T_{1}$ of H nuclei with a standard 
NMR technique called the 
inverse-recovery method to evaluate $\gamma_{\rm II} = 1/T_1$.
We will use $(\gamma_{\rm I},\gamma_{\rm II})=(0.41, 0.20)$~rad/s thus obtained, 
as listed in Ref.~\cite{Ho_2019}.

Now the dynamics of the total system including the environment 
is described by the GKLS 
equation~(\ref{eq:dyn2}) and our theoretical analysis developed 
in Sec.~\ref{sec:2} is straightforwardly
applicable to the molecular system.

\subsection{FID Signals}
\label{subsubsec:2}

\begin{figure}[h]
\begin{minipage}[t]{0.48\hsize}
\begin{center}
\includegraphics[width=60mm]{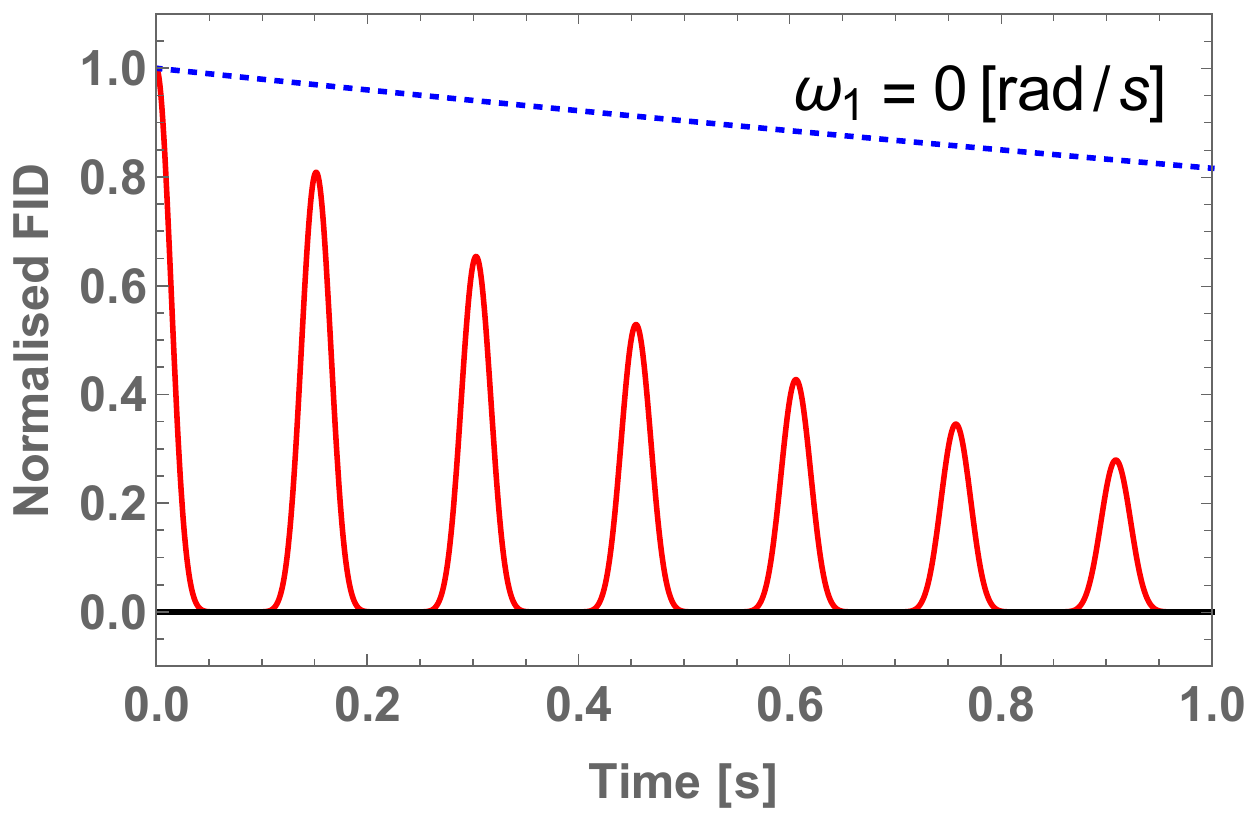}
\end{center}
\end{minipage}
\begin{minipage}[t]{0.48\hsize}
\begin{center}
\includegraphics[width=60mm]{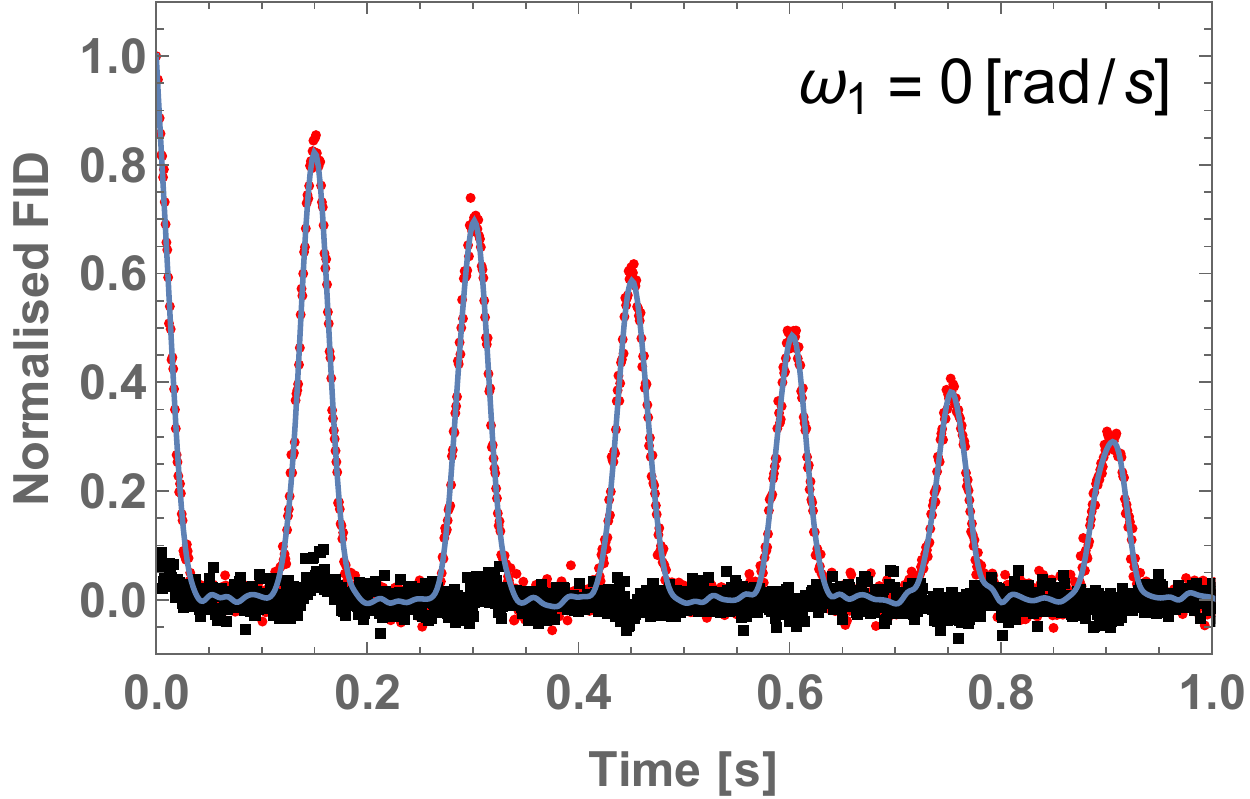}
\end{center}
\end{minipage}\\
\vspace{0.5cm}
\begin{minipage}[t]{0.48\hsize}
\begin{center}
\includegraphics[width=60mm]{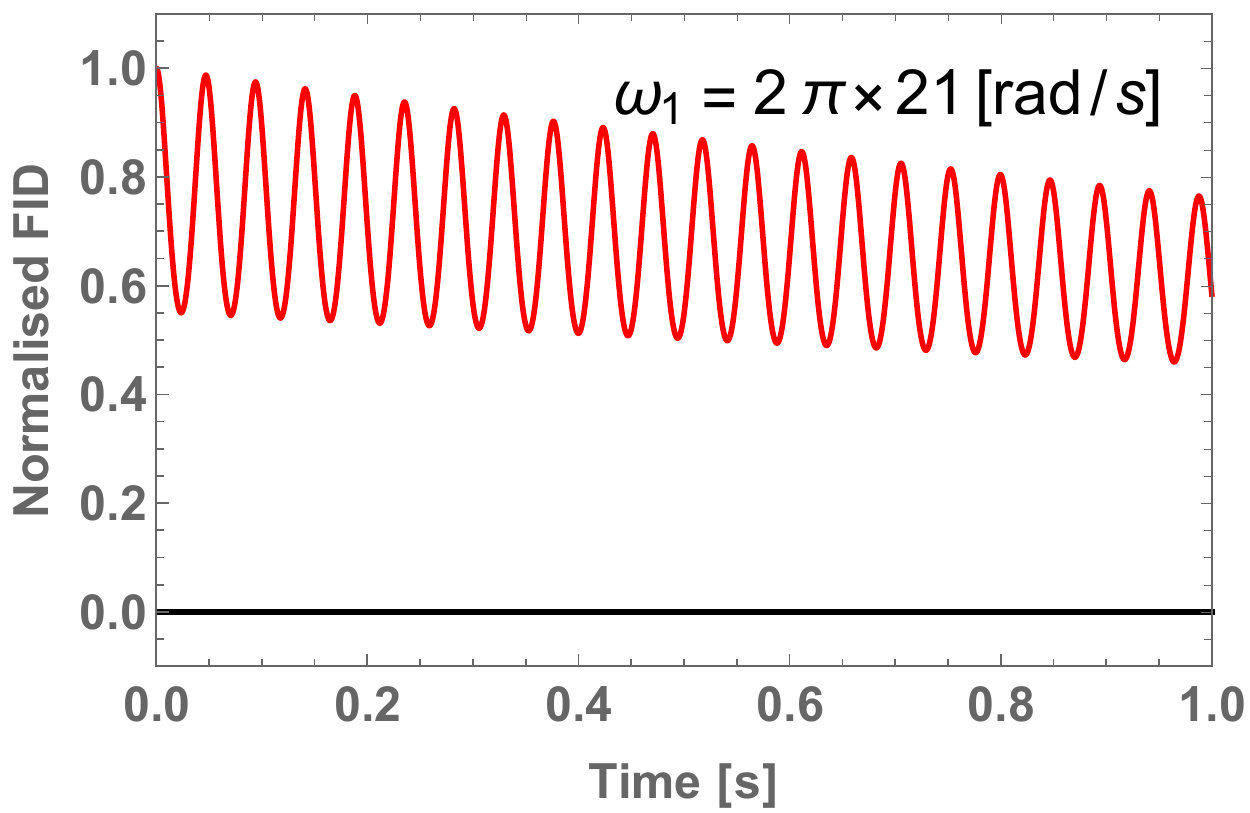}
\end{center}
\end{minipage}
\begin{minipage}[t]{0.48\hsize}
\begin{center}
\includegraphics[width=60mm]{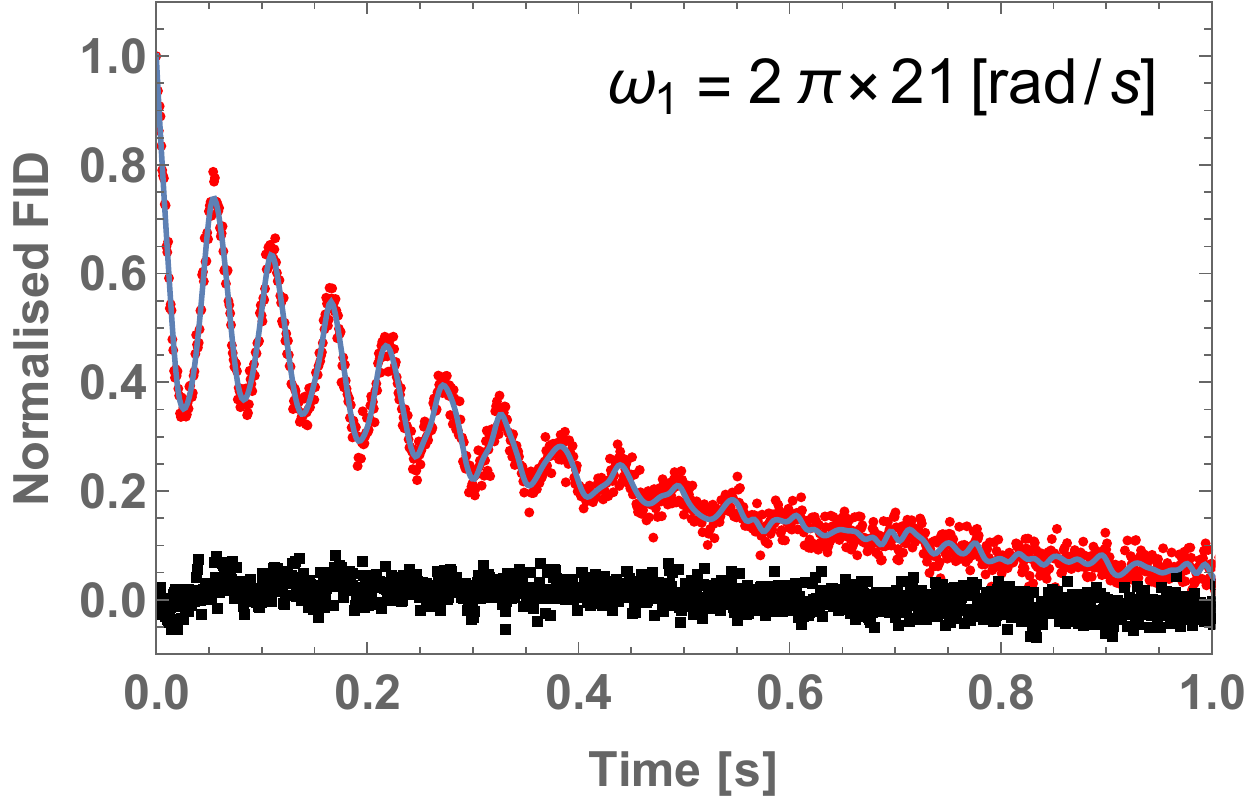}
\end{center}
\end{minipage}\\
\vspace{0.5cm}
\begin{minipage}[t]{0.48\hsize}
\begin{center}
\includegraphics[width=60mm]{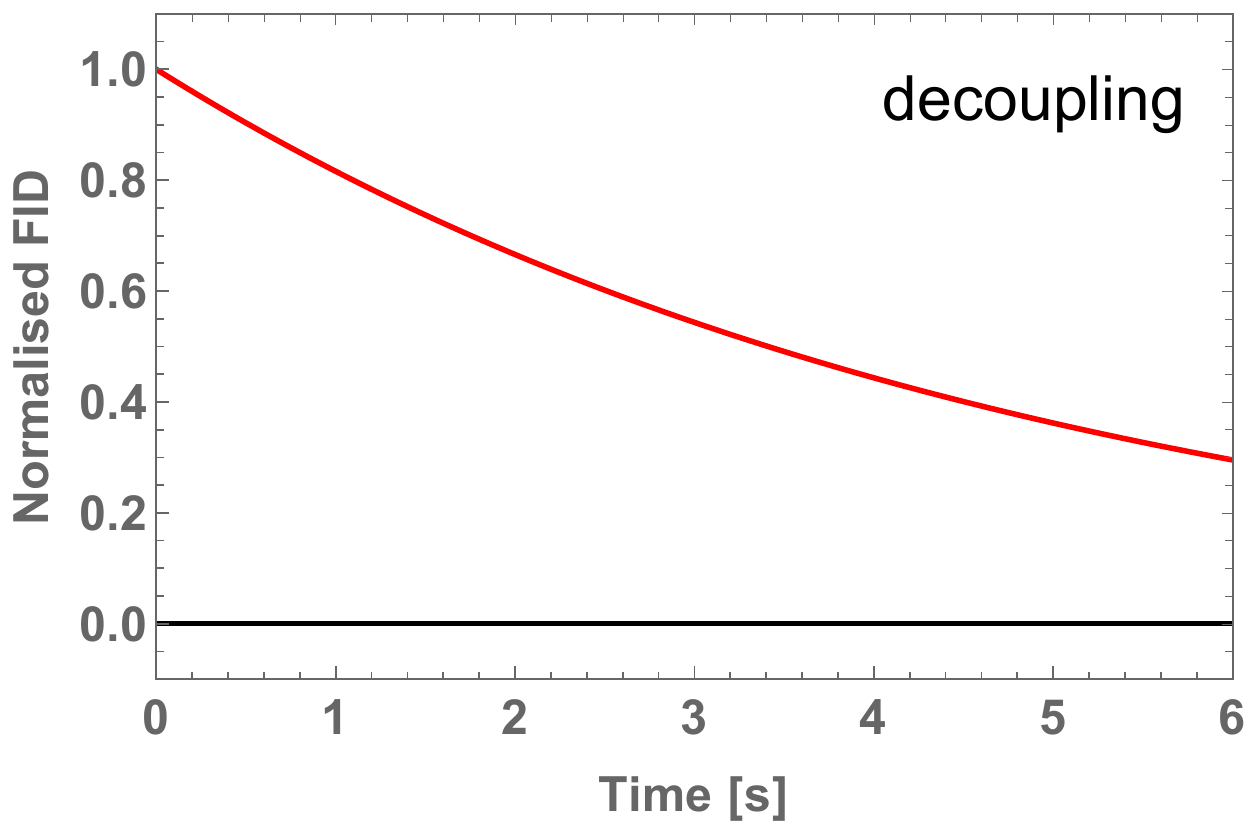}
\end{center}
\end{minipage}
\begin{minipage}[t]{0.48\hsize}
\begin{center}
\includegraphics[width=60mm]{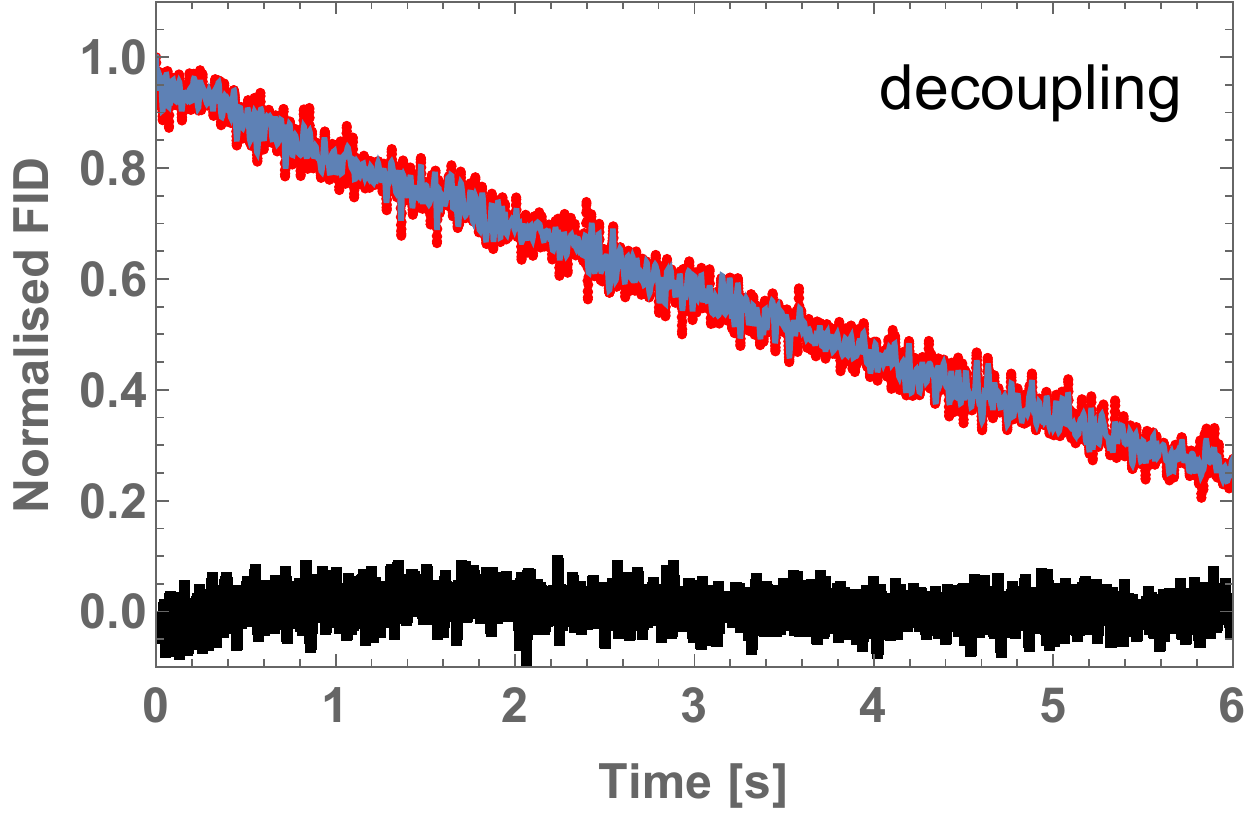}
\end{center}
\end{minipage}
\caption{Left three panels show the theoretical results 
while the right ones show the corresponding experimental data.
The red (black) lines represent the real (imaginary) parts of 
normalised FID's. In theoretical dynamics, we employed the parameters 
$(\gamma_{\rm I},\gamma_{\rm II},J)=(0.41, 0.20 ,2\pi \times 6.6)$~rad/s
~\cite{Ho_2019}.
$\omega_{1}$ is shown on the top right of each panel.
The blue lines in the right panels are obtained by moving-averaging 
the corresponding experimental data. 
The dotted line in the top-left panel is $e^{-\gamma_{I} t/2}$, which 
shows the direct influence of the Markovian environment on System~I. 
\label{fig:(1+12)}}
\end{figure}

Figure \ref{fig:(1+12)} shows the theoretical and experimental FID's.
The right panels are the normalized experimental FID's 
while the left panels show the theoretical ones. 
We plot the real (imaginary) parts of the normalised FID's.
In the right panels, we also show smooth curves obtained 
by moving-averaging the experimental data, which will be used 
to calculate non-Markovianity in the next subsection.

Clearly, theoretical calculations well reproduce the experimental FID's 
in both the Markovian (decoupling) and non-Markovian 
($\omega_{1}=0$~rad/s) limits, as discussed in Ref.~\cite{Ho_2019}.
The peaks in the top-left panel are smaller than $e^{-\gamma_{\rm I} t/2}$
which implies that the information stored in System~I can flow into 
the environment through System~II. In other words, the information 
can escape into the environment even if $\gamma_{\rm I} = 0$.
In the intermediate region ($\omega_{1}=2\pi\times 21$~rad/s case 
in Fig.~\ref{fig:(1+12)}), 
we can see that our theoretical dynamics qualitatively agree 
with the experimental data.
However, there is a quantitative difference between theory and 
experiment.
The observed decay is faster than the theoretical prediction.
This difference can be attributed to spatial inhomogeneity of 
$\omega_{1}$~\cite{Lapasar_2012}.
The sample was sealed in an NMR test tube with finite size and  $\omega_1$ is 
slightly different for TMS molecules at different parts of the tube. 
The observed FID signal
is a result of ensemble average over a macroscopic number of TMS molecules 
from various positions in the test tube and they have dynamics 
corresponding to the local $\omega_1$.
As a result, the observed FID signal involves average 
over $\omega_1$, namely average over
different dynamics, which leads to faster decay.
For the non-Markovian limit, such spatial inhomogeneity does not occur 
since  $\omega_{1} = 0$.
The Markovian (decoupling) limit is now achieved by WALTZ-16, 
a decoupling pulse sequence robust against spatial inhomogeneity of 
$\omega_{1}$ \cite{Claridge}.
This is the reason why theory well reproduces the experimental 
observation in both limits. 

\subsection{Engineering Non-Markovianity}

\begin{figure}[h]
\begin{center}
\includegraphics[width=160mm]{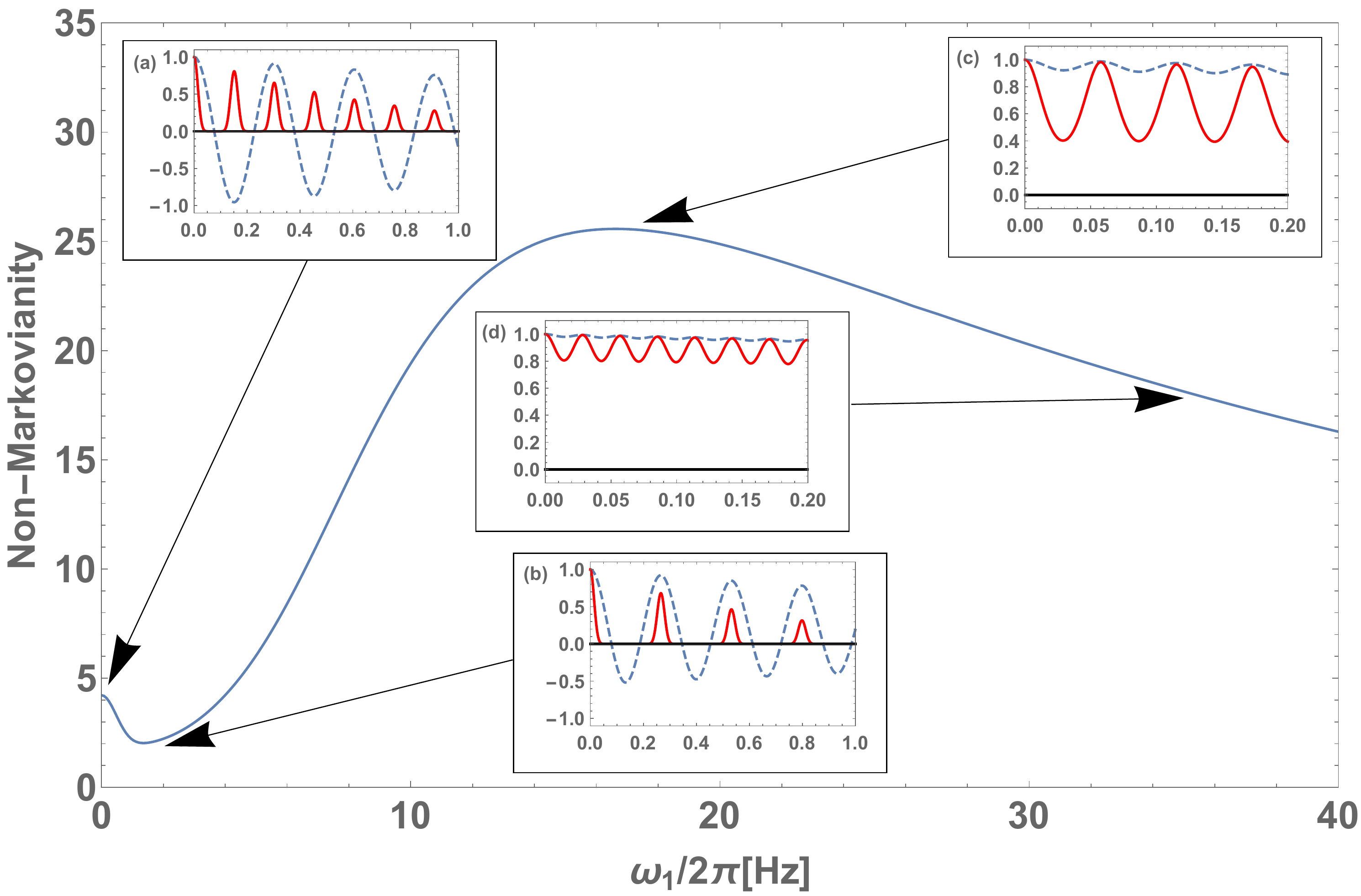}
\end{center}
\caption{
Non-Markovianity measure ${\cal N}$ as a function of $\omega_{1}$.
Each inset shows the dynamics of the FID signal $\beta_{12}(t)$ 
(red line) and $\beta_{1}(t)$ (dashed blue line) for the corresponding 
$\omega_{1}$:
(a) $\omega_{1}=0$~rad/s, (b) $\omega_{1}=2\pi\times1.8$~rad/s, 
(c) $\omega_{1}=2\pi\times17$~rad/s and (d) $\omega_{1}=2\pi\times35$~rad/s.
The plotted time interval of insets (a) and (b) is $[0,1.0]$~s 
while that of (c) and (d) is $[0,0.2]$~s. 
The parameters are $(\gamma_{\rm I}, \gamma_{\rm II}, J) = 
(0.41, 0.20, 2 \pi \times 6.6)$~rad/s. 
\label{fig:nonmarkovianity}
}
\end{figure}

Let us study how non-Markovianity measure $\cal N$ 
changes as a function of $\omega_1$
in our theory and experiment. 
We evaluate ${\cal N}$ from the analytical solution of $\beta_{12}(t)$ as shown in Fig.~\ref{fig:nonmarkovianity}.
The upper limit of the integration~(\ref{eq:nonmarkov2}) is taken to be 
$t=50~{\rm s} \sim 20/\gamma_{\rm I}$ instead of $t\rightarrow\infty$, 
which is sufficiently large compared to the time scale $2/\gamma_{\rm I}$ of the dynamics.

Note that  $\cal N$ does not decrease monotonically in this theoretical 
curve. There is a dip in the small $\omega_{1}$ region. This behaviour is 
understood
by examining insets in Fig. \ref{fig:nonmarkovianity}, which plots 
$\beta_{12}(t)$, the FID signal of the $(1+12)$-system.
We also plot $\beta_{1}(t)$ for comparison.
The magnitude of the signal is suppressed as a power of $n$ 
in the vicinity of  $t$
satisfying $\beta_{1}(t)=0$ (Inset (a) in Fig. \ref{fig:nonmarkovianity}).
The time intervals with such suppressed signals hardly contribute to 
${\cal N}$.
While $\omega_{1}$ increases, the oscillation centre of $\beta_{1}(t)$ 
is gradually lifted up.
Suppression occurs prominently when the lower end of the oscillation 
is located around zero (Inset (b));
thus ${\cal N}$ first decreases near $\omega_{1} \sim 0$ and 
hits the minimum.
After $\beta_{1}(t)$ is lifted up totally above zero, $n$ 
rather enhances the non-Markovianity since the oscillation 
is amplified according to the power of $n$ (Inset (c)).
This causes the dip shown in  Fig. \ref{fig:nonmarkovianity}.
In the remaining region, ${\cal N}$ monotonically decreases 
since the oscillation gradually disappears (Inset (d)).

\begin{figure}[h]
\begin{center}
\includegraphics[width=90mm]{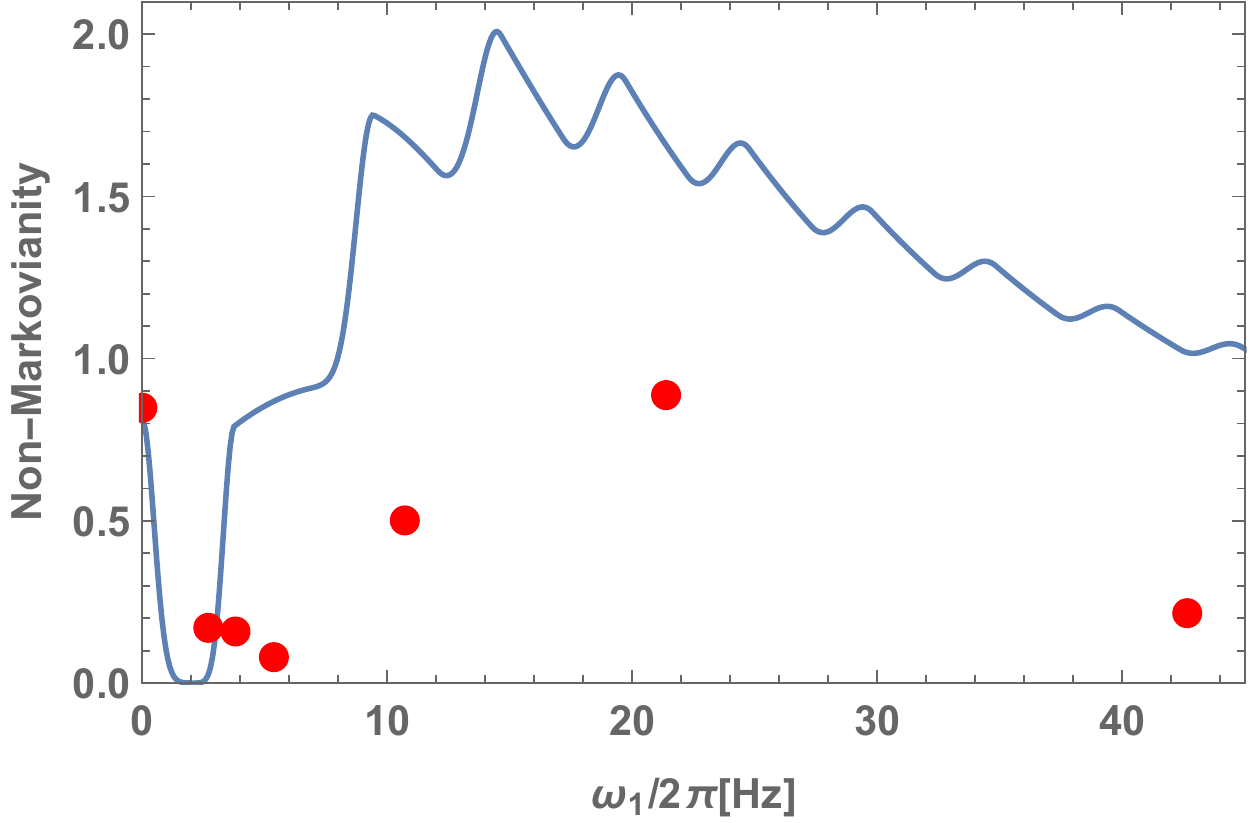}
\end{center}
\caption{Non-Markovianity ${\cal N} $with a truncated integral interval 
$[0,0.2]$~s.
The blue line is a theoretical result while the red points are obtained from the experimental data. 
\label{fig:nonmarexp}}
\end{figure}

We show ${\cal N}$ obtained from the experimental data 
in Fig.~\ref{fig:nonmarexp}. To avoid influence of the spatial 
inhomogeneity of $\omega_{1}$, we truncate the upper limit of 
the integration at a short time, $t=0.2$~s.
In experiments, ${\cal N}$ is evaluated by using the moving-averaged 
experimental data explained in Fig.~\ref{fig:(1+12)}.
We also give ${\cal N}$ for the theoretical dynamics with 
the same truncated integration interval to be compatible with 
the experimental results.
The outline of this theoretical curve is the same as the curve 
shown in Fig. \ref{fig:nonmarkovianity} although this curve has 
ripple structure because of the truncated integration interval.
We see that the dip and peak in the experimental result are 
qualitatively reproduced by the theoretical calculation.
The experimental results, however, basically shows smaller ${\cal N}$ 
than theoretical ones.
This difference can be understood as the effect of the spatial 
inhomogeneity of $\omega_{1}$ as discussed in Sec.~\ref{subsubsec:2}.
The faster decay of FID signals in the experiment makes ${\cal N}$ smaller.

\section{Summary}
\label{sec:4}

We have proposed an open-system model of which dynamics can be continuously 
tuned from non-Markovian to Markovian by changing an external field.
The model consists of System~I that is the principal system of interest, 
System~II surrounding System~I, and Markovian environment.
We have shown that the dynamics of this model can be solved analytically 
with a reasonable initial condition.
We compared our theoretical results with the experimental data.

We have shown that the results of the theoretical model qualitatively 
agree with the experimental results:
in particular, the transition from Markovian to non-Markovian 
dynamics is well reproduced.
Then we evaluated non-Markovianity of our model by introducing 
a non-Markovianity measure $\cal N$ based on the trace distance.
Our model is expected to serve to understand non-Markovian open systems.

\section*{Acknowledgement}

SK and YK would like to thank CREST (JPMJCR1774) JST. MN is partly supported 
by JSPS Grants-in-Aid for Scientific Research (Grant Number 20K03795).

\appendix
\section{Exact solution of Eq.~(\ref{eq:key})}

Here we will show the exact form of $b_{0}(t)$ by solving Eq.~(\ref{eq:key}).
To do this, it is enough to find the eigenvalues and eigenvectors of 
${\bold M}'$ defined as
\begin{equation}
{\bold M}':= 2 {\bold M}^{\rm T}=
\begin{pmatrix}
0&0&-i J\\
0&-\gamma_{\rm II}&-2\omega_{1}\\
-i J&2\omega_{1}&-2\gamma_{\rm II}
\end{pmatrix}
.
\end{equation}
The eigenvalues of ${\bold M}'$ are
\begin{align}
\lambda_{1}=-\gamma_{\rm II}-\frac{D}{C}+\frac{C}{3},\,\,\,\, 
\lambda_{2}=-\gamma_{\rm II}-\frac{D}{C}
-\Bigl(\frac{1-\sqrt{3}i}{2}\Bigr)\frac{C}{3}, \,\,\,\, 
\lambda_{3}&= \lambda_{2}^* \nonumber \\
C =\Bigl( 54\gamma_{\rm II}\omega_{1}^{2}+3\sqrt{3}
\sqrt{108\gamma_{\rm II}^{2}\omega_{1}^{4}+D^{3}}\Bigr)^{\frac{1}{3}}, 
\,\,\,\, 
D = J^{2}-\gamma_{\rm II}^{2}+4\omega_{1}^{2}
\end{align}
Note that $C$ is always real with our parameters 
$(\gamma_{\rm I},\gamma_{\rm II},J)=(0.41,0.20,2\pi\times6.6)$~rad/s.
The corresponding (unnormalised) eigenvectors are given as
\begin{equation}
\vec{v}_{1}=
\begin{pmatrix}
1\\
-2i\frac{\omega_{1}}{J}\frac{\lambda_{1}}{\lambda_{1}+\gamma_{\rm II}}\\
i\frac{\lambda_{1}}{J}
\end{pmatrix}
,
\vec{v}_{2}=
\begin{pmatrix}
1\\
-2i\frac{\omega_{1}}{J}\frac{\lambda_{2}}{\lambda_{2}+\gamma_{\rm II}}\\
i\frac{\lambda_{2}}{J}
\end{pmatrix}
,
\vec{v}_{3}=
\begin{pmatrix}
1\\
-2i\frac{\omega_{1}}{J}\frac{\lambda^{*}_{2}}{\lambda^{*}_{2}+\gamma_{\rm II}}\\
i\frac{\lambda^{*}_{2}}{J}
\end{pmatrix}
.
\end{equation}
By using the above eigenvalues and eigenvectors, the solution is written as
\begin{equation}
\begin{pmatrix}
b_{0}\\
b_{y}\\
b_{z}
\end{pmatrix}
=\sum_{i=1,2,3}u_{i}\vec{v}_{i}\exp(\lambda_{i}t/2)
\end{equation}
where $\{u_{i}\}_{i=1,2,3}$ are constant parameters determined 
by the initial condition.
When assigning the initial values $(b_{0}(0),b_{y}(0),b_{z}(0))=(1,0,0)$, 
we obtain
\begin{align}
u_{1}&=\frac{|\lambda_{2}|^{2}(\lambda_{1}+\gamma_{\rm II})}
{\gamma_{\rm II}\bigl((\lambda_{1}-\lambda^{R}_{2})^{2}
+(\lambda_{2}^{I})^{2}\bigr)},\nonumber\\
u_{2}&=
\frac{\lambda^{2}_{1}\gamma_{\rm II}-\lambda_{1}(|\lambda_{2}|^{2}
+2\lambda^{R}_{2}\gamma_{\rm II})}
{2\gamma_{\rm II}\bigl((\lambda_{1}-\lambda^{R}_{2})^{2}
+(\lambda_{2}^{I})^{2}\bigr)}
+i\frac{\lambda_{1}(\lambda_{1}-\lambda^{R}_{2})(|\lambda_{2}|^{2}
+\lambda^{R}_{2}\gamma_{\rm II})+\lambda^{1}(\lambda^{I}_{2})^{2}
\gamma_{\rm II}}{2\lambda^{I}_{2}\gamma_{\rm II}
\bigl((\lambda_{1}-\lambda^{R}_{2})^{2}+(\lambda_{2}^{I})^{2}\bigr)},
\nonumber\\
u_{3}&=u_2^*,
\end{align}
where we introduce the real (imaginary) part of 
$\lambda_{2}$: $\lambda_{2}=\lambda^{R}_{2}
+i\lambda^{I}_{2},~\lambda^{R}_{2},\lambda^{I}_{2}\in {\mathbb R}$.
Thus, the explicit form of $b_{0}(t)$ is
\begin{align}
b_{0}(t)&=u_{1}\exp(\lambda_{1}t/2)
+u_{2}\exp(\lambda_{2}t/2)+u_{3}\exp(\lambda_{3}t/2)
\nonumber \\
&=u_{1}\exp(\lambda_{1}t/2)+2\exp(\lambda^{R}_{2}t/2)
\bigl(u^{R}_{2}\cos(\lambda^{I}_{2}t/2)
-u^{I}_{2}\sin(\lambda^{I}_{2}t/2)\bigr),
\end{align}
where $u_2^R$ $(u_2^I)$ is a real (imaginary) part of $u_2$. 
Note that $b_0(t)$ is always real with our parameters.


\end{document}